\documentclass[preprint,12pt,authoryear]{elsarticle}

\usepackage[utf8]{inputenc}
\usepackage[T1]{fontenc}
\usepackage{lmodern}
\usepackage{amsmath,amssymb,mathtools}
\usepackage{graphicx}
\usepackage{booktabs}
\usepackage{threeparttable}
\usepackage{multirow}
\usepackage{longtable}
\usepackage{array}
\usepackage{siunitx}
\usepackage{xcolor}
\usepackage{hyperref}
\usepackage{lineno}
\usepackage{microtype}
\usepackage{enumitem}
\usepackage{subcaption}
\usepackage{float}
\usepackage{pdflscape}
\usepackage{caption}
\usepackage{geometry}
\hypersetup{colorlinks=true,linkcolor=blue!55!black,citecolor=blue!55!black,urlcolor=blue!55!black}
\journal{Journal of Hydrology: Regional Studies (working draft)}
\biboptions{authoryear,round}
\newcommand{\mthree}{\mathrm{m^3\,s^{-1}}}
\newcommand{\WASI}{\mathrm{WASI}}

\begin{document}
\begin{frontmatter}

\title{From Daily Fluctuations to Annual Hydrological Cycles: A Wavelet-Based Analysis of Nonstationary Seasonality in Senegal River Hydropower Inflows}

\author[lmdan,uon]{Steeven B. Affognon\corref{cor1}}
\ead{belvinos@gmail.com}
\author[lmdan]{Babacar M. Ndiaye}
\author[lmdan]{Pierre Mendy}
\author[esp,uqtr]{Cheikh M. F. Kebe}
\cortext[cor1]{Corresponding author.}

\address[lmdan]{Laboratory Mathematics of Decision and Numerical Analysis, Cheikh Anta Diop University of Dakar, Dakar, Senegal}
\address[uon]{Department of Mathematics, University of Nairobi, Nairobi, Kenya}
\address[esp]{Laboratoire Eau, Energie, Environnement et procédés Industriels (LE3PI), UCAD, Dakar, Senegal}
\address[uqtr]{Centre de Tests des Systèmes Solaires (CT2S), Dakar, Senegal}

\begin{abstract}
Reliable representation of inflow seasonality is essential for hydropower scheduling, stochastic scenario construction, and seasonal-storage valuation. However, a fixed monthly climatology cannot determine whether the timing, amplitude, and persistence of the annual hydrological cycle change through time. This study develops a reproducible Fourier--wavelet framework for analysing daily inflows at Bafing Makana, F\'elou, and Gouina in the Senegal River system. The dataset comprises 65,631 daily observations, with 21,877 observations per site, spanning nearly 60 years from 1 January 1961 to 23 November 2020. Harmonic regression, Welch spectral analysis, stationary wavelet multiresolution analysis, Morlet continuous wavelet transforms, directional seasonality metrics, red-noise significance testing, non-parametric trend and change-point tests, and cross-wavelet coherence are combined within a single analytical workflow. Because a daily dyadic detail \(D_j\) approximately represents periods in the interval \([2^j,2^{j+1}]\) days, the annual cycle is associated primarily with \(D_8\), corresponding to approximately 256--512 days, rather than with a level-4 approximation. Under a validation score comparing the reconstructed annual-scale component with the Fourier seasonal benchmark, the Haar wavelet was selected among Haar, Daubechies-4, Symlet-4, and Coiflet-2. The \(D_8\) component accounted for 18.85--19.89\% of total stationary-wavelet-transform energy, while the selected Fourier harmonic models explained 73.3--78.7\% of daily inflow variability. Mean Wavelet Annual Seasonality Index values ranged from 0.538 to 0.584, climatological maxima occurred in September, and mean reconstructed annual-component peaks occurred near days 252--254. Mean annual-band wavelet coherence ranged from 0.988 to 0.999, demonstrating an exceptionally synchronized annual cycle across the three sites. No robust monotonic trend was detected in annual-seasonality strength, reconstructed annual amplitude, or peak timing after multiplicity correction. However, Pettitt tests identified statistically significant changes in annual-component amplitude or mean inflow during approximately 1971--1976. These shifts are consistent with a regional regime-transition interpretation, although attribution requires independent rainfall, reservoir-operation, and gauge-history information. The proposed framework provides a statistically transparent basis for scale-aware, multivariate inflow scenario generation within MOSSHOOS--Plan4RES and is transferable to other data-scarce hydropower systems.
\end{abstract}

\begin{keyword}
Senegal River \sep hydropower inflow \sep seasonality \sep stationary wavelet transform \sep continuous wavelet transform \sep Fourier analysis \sep wavelet coherence \sep change point \sep stochastic scenarios
\end{keyword}
\end{frontmatter}

\section{Introduction}
Hydropower planning depends not only on the average volume of water entering a reservoir but also on the timing, persistence, and predictability of that inflow. In strongly seasonal basins, errors in the estimated onset, peak, or recession of the annual cycle can propagate into storage targets, firm-energy assessments, spill risk, and shortage exposure. These concerns are particularly important in West African transboundary systems, where hydroclimatic variability, sparse observations, infrastructure development, and competing water demands interact over multiple time scales \citep{Bodian2016,Bodian2018,Bodian2020,Wilcox2018,Bruckmann2022,Ndiaye2023,Sambou2023,Goudiaby2024}.

Conventional seasonal analysis generally relies on monthly climatologies, harmonic regression, or stationary autoregressive models. Fourier analysis is valuable for estimating smooth periodic components and global spectral peaks, but its trigonometric basis functions extend across the entire record and therefore cannot localize transient changes in seasonal strength \citep{Welch1967,EmeryThomson2001}. Hydrological signals are often intermittent, skewed, nonstationary, and affected by abrupt wet and dry episodes. Wavelet methods address this limitation by decomposing a time series simultaneously by time and scale \citep{Mallat1989,Daubechies1992,PercivalWalden2000,Labat2005,Sang2013,Addison2017}. Applications have included rainfall--runoff separation, streamflow variability, climate teleconnections, sediment dynamics, droughts, and reservoir regime changes \citep{Nakken1999,CoulibalyBurn2004,Carey2013,Su2019,Juez2021,Juez2022,Juez2023,HadiTombul2018,Okkan2012}.

Two methodological problems remain important. First, dyadic discrete wavelet levels must be interpreted according to the sampling interval. For daily data, level 4 is a submonthly scale; it cannot be labelled annual seasonality. Since $2^8=256<365.25<512=2^9$, the annual cycle lies primarily in the $D_8$ band. Second, visually prominent wavelet power is not necessarily statistically meaningful. Edge effects, red-noise persistence, smoothing choices, and multiple-scale dependence can generate misleading features unless the cone of influence and significance testing are respected \citep{TorrenceCompo1998,MaraunKurths2004,MaraunEtAl2007,Liu2007,Grinsted2004,RodriguezMurillo2020,HuSi2016,HuSi2021}.

The Senegal River literature documents pronounced hydroclimatic variability and regime changes, including shifts near the major Sahel drought period and subsequent partial recovery \citep{Diop2018,Wilcox2018,Bodian2020,Bruckmann2022,Ndiaye2023}. However, a site-by-site, daily, scale-resolved comparison of the hydropower inflows at Bafing Makana, F\'elou, and Gouina has not been established in the supplied project material. This study therefore asks four questions: (i) which scales dominate daily inflow variability; (ii) how stable are annual amplitude and timing; (iii) how synchronized are the three sites at annual and interannual scales; and (iv) how can the resulting decomposition support stochastic scenario generation for seasonal storage optimization?

The contributions are: a validated distinction between Fourier seasonality and dyadic wavelet bands; a Wavelet Annual Seasonality Index (WASI); scale-specific energy and annual-component diagnostics; time-varying Morlet power with red-noise significance; cross-site wavelet coherence; non-parametric trend and change-point tests; and a fully reproducible Python/Colab package. The analysis does not use wavelets merely to remove variability. Instead, it separates persistent seasonal structure from physically meaningful detail components that should be retained in stochastic inflow scenarios.

\section{Study area and data}
\subsection{Senegal River hydropower sites}
\begin{figure}[H]
\centering
\includegraphics[width=0.85\textwidth]{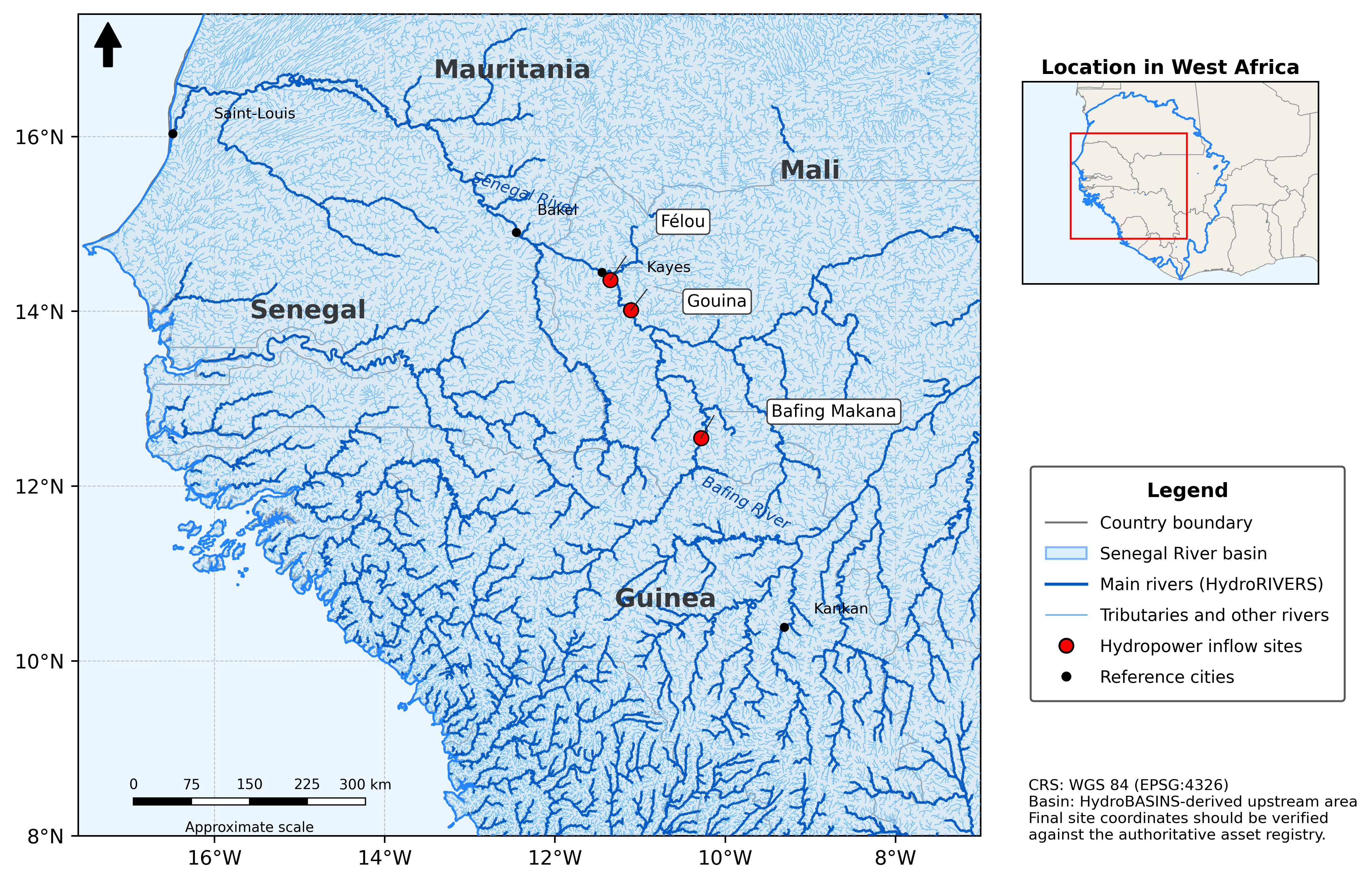}
\caption{Geographical context of Bafing Makana, F\'elou, and Gouina within the Senegal River system. The basin boundary is generated from HydroBASINS, the river network from HydroRIVERS/Natural Earth sources, and country boundaries from Natural Earth using the accompanying Python map script. The map is used for spatial orientation only; all numerical results are derived from the dated inflow records.}
\label{fig:map}
\end{figure}

The investigated sites are Bafing Makana, F\'elou, and Gouina (Fig.~\ref{fig:map}). Bafing Makana represents the upper Bafing branch, while F\'elou and Gouina are closely connected downstream sites. The upper basin is influenced by the West African monsoon and exhibits a marked wet-season pulse, with implications for hydropower production and transboundary water management \citep{Bodian2016,Bodian2020,Sambou2023}. Fig.~\ref{fig:map} is used only for spatial orientation; the statistical analyses do not use site coordinates. The site positions shown in the figure should therefore be interpreted as contextual map locations rather than inputs to the Fourier, wavelet, trend, or coherence calculations.

\subsection{Daily inflow data and quality control}
The input file contains date, plant, scenario, and inflow in $\mthree$. Each site has 21,877 records from 1961-01-01 to 2020-11-23, yielding 65,631 observations. The calendar is complete, with no duplicated or missing dates in the regularized series. The file does not contain observation-status codes, gauge-quality flags, rating-curve versions, or a separate field distinguishing physical zero flow from censored, rounded, operational, or missing-coded values. Recorded non-positive values number 182 for Bafing Makana, 103 for F\'elou, and 1,373 for Gouina. These observations were retained in the principal analysis as recorded values; they were not reclassified because no quality-flag information in the input file supports such a decision. Consequently, interpretations of low-flow behaviour remain conditional on the observational status of these records. The main seasonal-scale conclusions are less directly affected by zero coding because they rely on annual timing, scale energy, Fourier spectra, and wavelet coherence rather than on the lower tail alone.

\begin{figure}[H]
\centering
\includegraphics[width=0.98\textwidth]{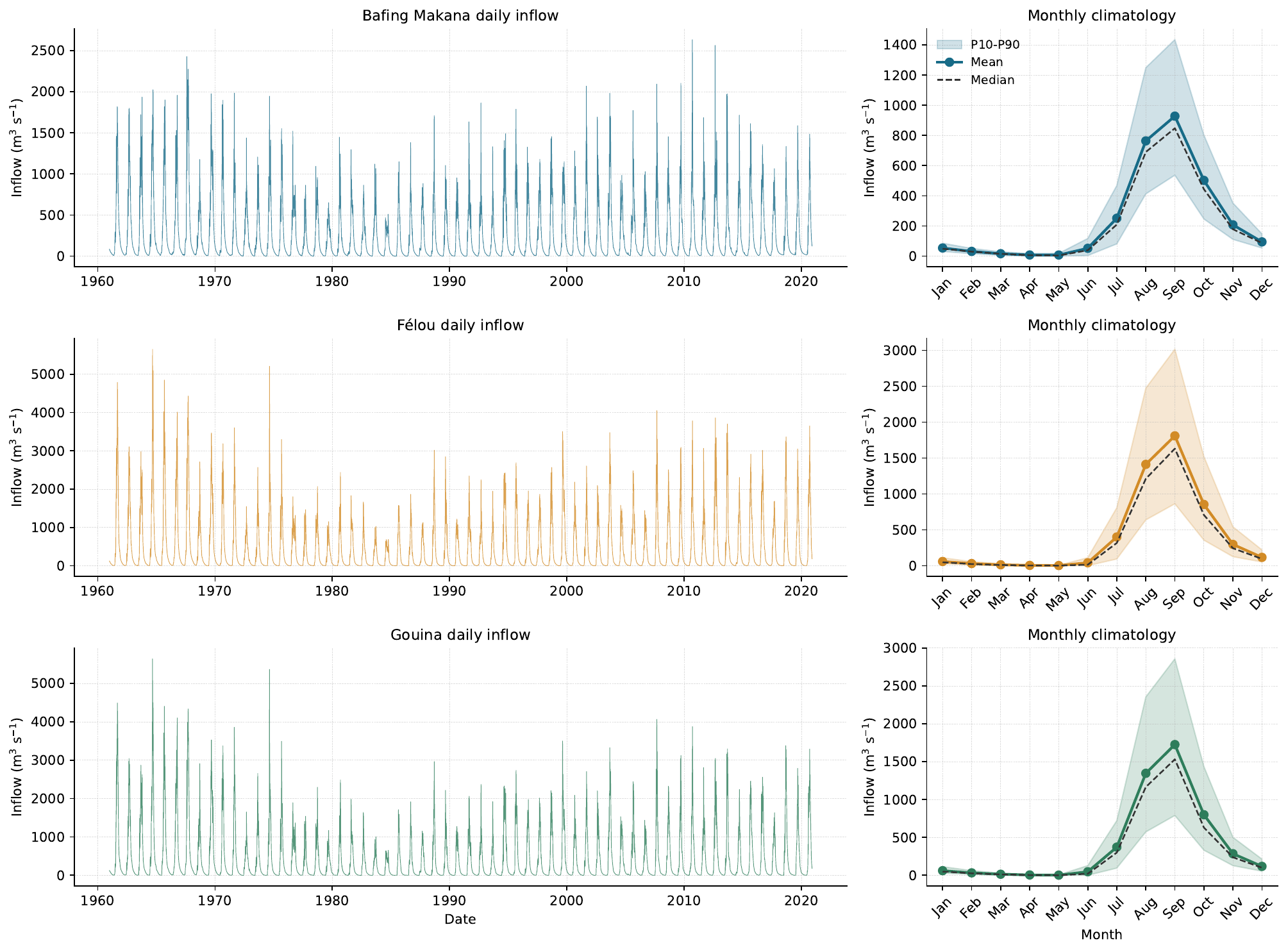}
\caption{Daily inflows and monthly climatology for the three hydropower inflow series. The upper panels show the complete daily records, while the lower panels summarize the average annual cycle; shaded envelopes denote the empirical P10--P90 range and highlight interannual variability around the monthly climatological mean.}
\label{fig:daily}
\end{figure}

The series are strongly right-skewed (Fig.~\ref{fig:daily}). Means are 244.15, 421.37, and 402.32 $\mthree$ for Bafing Makana, F\'elou, and Gouina, respectively, while coefficients of variation range from 1.44 to 1.67 (Table~\ref{tab:desc}). All monthly climatologies peak in September.

\begin{table}[H]
\centering
\caption{Descriptive statistics of the daily inflow records. The table reports sample size, mean, standard deviation, coefficient of variation, median, 90th percentile, and maximum discharge for each site.}
\label{tab:desc}
\small
\begin{tabular}{lrrrrrrr}
\toprule
Site & $n$ & Mean & SD & CV & Median & P90 & Maximum\\
 & & \multicolumn{6}{c}{$\mthree$ except CV}\\
\midrule
Bafing Makana & 21,877 & 244.15 & 351.92 & 1.441 & 74.79 & 761.74 & 2,634.00\\
F\'elou & 21,877 & 421.37 & 700.77 & 1.663 & 79.70 & 1,361.06 & 5,655.36\\
Gouina & 21,877 & 402.32 & 672.37 & 1.671 & 85.23 & 1,306.00 & 5,641.00\\
\bottomrule
\end{tabular}
\end{table}

\section{Methods}
\subsection{Monthly climatology and directional seasonality}
For site $p$ and calendar month $m$, the monthly climatological mean is
\begin{equation}
\bar Q_{p,m}=\frac{1}{N_m}\sum_{t:\,m(t)=m}Q_p(t).
\end{equation}
Annual timing was also estimated using directional statistics. For day-of-year angle $\theta_t=2\pi(d_t-1)/L_y$ and positive weight $w_t=\max(Q_t,0)$,
\begin{equation}
C_y=\sum_{t\in y} w_t\cos\theta_t,\qquad
S_y=\sum_{t\in y} w_t\sin\theta_t,
\end{equation}
with timing $\phi_y=\operatorname{atan2}(S_y,C_y)$ and concentration
\begin{equation}
R_y=\frac{\sqrt{C_y^2+S_y^2}}{\sum_{t\in y}w_t}.
\end{equation}
Directional statistics avoid the discontinuity between December and January and have recently been advocated for global streamflow-seasonality comparisons \citep{Berghuijs2025}.

\subsection{Fourier harmonic baseline and Welch spectrum}
The smooth seasonal baseline was represented by
\begin{equation}
\theta_p(t)=a_{0,p}+\sum_{k=1}^{K_p}\left[a_{k,p}\cos\left(\frac{2\pi kt}{365.25}\right)+b_{k,p}\sin\left(\frac{2\pi kt}{365.25}\right)\right].
\label{eq:fourier}
\end{equation}
Candidate $K=1,\ldots,8$ models were fitted by least squares and selected by the Bayesian information criterion. Welch periodograms provided a complementary global spectral view \citep{Welch1967}. Fourier analysis serves as a transparent periodic benchmark, not as the final nonstationary model.

\subsection{Stationary wavelet multiresolution analysis}
For a mother wavelet $\psi$ and scaling function $\phi$, the signal is decomposed as
\begin{equation}
Q_p(t)=A_{J,p}(t)+\sum_{j=1}^{J}D_{j,p}(t),
\label{eq:mra}
\end{equation}
where $A_J$ is the approximation and $D_j$ the detail at scale $j$ \citep{Haar1910,Mallat1989,Daubechies1992,PercivalWalden2000}. We used the stationary wavelet transform (SWT) because it is translation invariant and returns a component at every time point. Candidate families were Haar, Daubechies-4, Symlet-4, and Coiflet-2, with $J=10$.

For daily data, $D_j$ approximately covers $[2^j,2^{j+1}]$ days. Therefore, $D_1$ reflects 2--4-day fluctuations, $D_4$ 16--32-day variability, $D_7$ 128--256-day variability, and $D_8$ the 256--512-day band containing 365.25 days. The annual component was consequently defined as $D_8$, while $A_{10}$ represents variability slower than roughly 1,024 days.

Two related annual-period definitions are used because the discrete and continuous wavelet analyses have different resolutions. In the SWT, the dyadic $D_8$ component is a broad discrete band, approximately 256--512 days, and is therefore the appropriate reconstructed annual-scale component. In the CWT and coherence analyses, periods are sampled more densely, so annual power and synchronization are summarized over the narrower 300--450-day window centred on the hydrological year. This window lies inside the $D_8$ dyadic interval and reduces contamination from the shorter $D_7$ band and the longer $D_9$ band. Thus, $D_8$ is used for discrete multiresolution reconstruction, whereas 300--450 days is used for continuous annual-band summaries.

Wavelet $w$ was selected using a validation score comparing the scaled $D_8$ component to the Fourier fit:
\begin{equation}
\mathcal S(w)=\frac{\mathrm{RMSE}\{\theta_p,\widehat a+\widehat bD_{8,p}^{(w)}\}}{\mathrm{SD}(\theta_p)}+\left[1-\max\{\mathrm{Cor}(\theta_p,D_{8,p}^{(w)}),0\}\right].
\label{eq:score}
\end{equation}
This selection is conditional on the stated score and does not imply that Haar is universally optimal for hydrological applications.

Scale energy was computed as
\begin{equation}
E_{j,p}=\sum_t D_{j,p}^2(t),\qquad
\pi_{j,p}=100\frac{E_{j,p}}{E_{A_J,p}+\sum_{r=1}^{J}E_{r,p}}.
\label{eq:energy}
\end{equation}

\subsection{Continuous wavelet transform and significance}
The complex Morlet CWT is
\begin{equation}
W_p(\tau,s)=\frac{1}{\sqrt{s}}\int Q_p(t)\psi^*\!\left(\frac{t-\tau}{s}\right)\,dt,
\end{equation}
with local power $P_p(\tau,s)=|W_p(\tau,s)|^2$ and global spectrum $\bar P_p(s)=N^{-1}\sum_\tau P_p(\tau,s)$ \citep{TorrenceCompo1998}. Scales covered periods from 2 to 2,048 days with 12 sub-octaves. Power was tested against an AR(1) red-noise null at 95\% significance. The cone of influence (COI) marks regions in which boundary effects are material \citep{MaraunKurths2004,MaraunEtAl2007,Liu2007,RodriguezMurillo2020}.

\subsection{Wavelet Annual Seasonality Index}
For year $y$, the Wavelet Annual Seasonality Index is defined as the scale-normalized annual-band energy fraction
\begin{equation}
\WASI_p(y)=
\frac{\displaystyle\sum_{\tau\in y}\sum_{s\in[300,450]}P_p(\tau,s)/s}
{\displaystyle\sum_{\tau\in y}\sum_{s\in[2,2048]}P_p(\tau,s)/s}.
\label{eq:wasi}
\end{equation}
Only years with sufficient valid annual-band support outside severe edge influence were retained. The $1/s$ normalization reduces the mechanical increase of raw wavelet power with scale \citep{Liu2007}.

\subsection{Cross-wavelet coherence}
For sites $p$ and $r$, the cross-wavelet transform is $W_{pr}=W_pW_r^*$. Squared coherence is
\begin{equation}
R_{pr}^{2}(\tau,s)=
\frac{\left|S\{s^{-1}W_{pr}(\tau,s)\}\right|^2}
{S\{s^{-1}|W_p(\tau,s)|^2\}\,S\{s^{-1}|W_r(\tau,s)|^2\}},
\label{eq:coherence}
\end{equation}
where $S$ denotes time--scale smoothing \citep{Grinsted2004,Carey2013,HuSi2016,HuSi2021}. To reduce computation and emphasize management-relevant scales, coherence was calculated from weekly means. Average coherence was summarized for the annual band (300--450 days) and an interannual band (730--2,048 days).

\subsection{Trend, slope, and change-point tests}
Monotonic trends were assessed using Kendall's $\tau$ and Sen's median slope \citep{Mann1945,Kendall1975,Sen1968,Helsel2020}. Abrupt distributional shifts were assessed with the Pettitt test \citep{Pettitt1979}. These non-parametric tests are appropriate for skewed hydrological metrics but do not establish causality. Because multiple site--metric tests were conducted, unadjusted $p$-values were complemented by Bonferroni correction across the 15 Kendall trend tests. Benjamini--Hochberg false-discovery-rate adjustment was also examined as a sensitivity check \citep{BenjaminiHochberg1995}. Results are interpreted as robust only when they remain significant after multiplicity correction or are supported consistently across related diagnostics.

\subsection{Software and reproducibility}
The workflow was implemented in Python with NumPy, SciPy, Matplotlib, PyWavelets, pandas, GeoPandas, and pycwt \citep{Harris2020,Virtanen2020,Hunter2007,Lee2019}. The reproducibility package contains the full script, Colab notebook, environment file, intermediate CSV files, and all publication figures. Redistribution of the raw daily inflow file is governed by the data-owner conditions stated in the Data and code availability section. Block bootstrap and clustering modules are not used to alter the present results, but are included in the broader MOSSHOOS workflow for later scenario generation \citep{Kunsch1989,PolitisRomano1994,Lloyd1982}.

\section{Results}
\subsection{Observed seasonal structure}
The three inflow series exhibit a highly concentrated unimodal annual cycle (Fig.~\ref{fig:daily}). September climatological means are 927.81 $\mthree$ at Bafing Makana, 1,808.33 $\mthree$ at F\'elou, and 1,725.42 $\mthree$ at Gouina. Mean directional timing occurs near day 255--256, with concentration $R$ between 0.754 and 0.813. These diagnostics confirm a strong shared wet-season pulse but also substantial interannual amplitude variability.

\subsection{Fourier seasonality and global spectra}
\begin{figure}[H]
\centering
\includegraphics[width=0.98\textwidth]{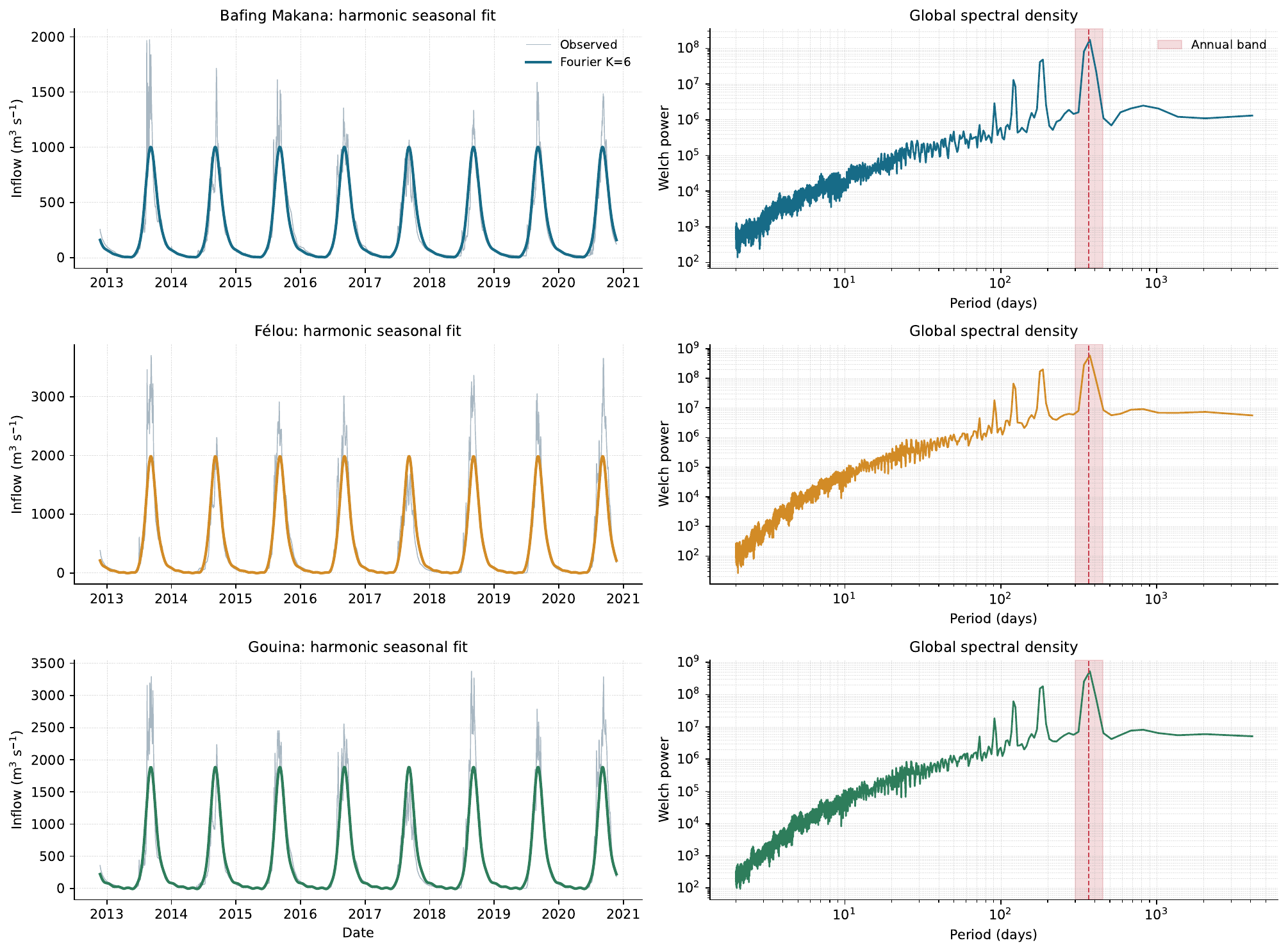}
\caption{Fourier seasonal fits and Welch power spectra for the three daily inflow series. The fitted curves represent the average periodic structure estimated over the complete observation period, while the spectra show how signal power is distributed across temporal periods. The shaded red band identifies the 300--450-day interval used to represent the annual-period neighbourhood in continuous-wavelet summaries.}
\label{fig:fourier}
\end{figure}

BIC selected 6 harmonics for Bafing Makana, 7 for F\'elou, and 6 for Gouina (Table~\ref{tab:fourier}). The corresponding $R^2$ values are 0.787, 0.748, and 0.733. Thus, a smooth annual harmonic model explains most but not all daily variability, leaving substantial non-Gaussian, short-scale, and interannual structure. Welch spectra display a dominant annual neighbourhood and lower-frequency variability (Fig.~\ref{fig:fourier}).

\begin{table}[H]
\centering
\caption{Fourier harmonic models selected by the Bayesian information criterion for each inflow series. The table reports the selected number of harmonics, the proportion of daily inflow variability explained by the fitted seasonal component, and the corresponding root-mean-square error.}
\label{tab:fourier}
\begin{tabular}{lrrr}
\toprule
Site & Harmonics & $R^2$ & RMSE ($\mthree$)\\
\midrule
Bafing Makana & 6 & 0.787 & 162.56\\
F\'elou & 7 & 0.748 & 351.93\\
Gouina & 6 & 0.733 & 347.48\\
\bottomrule
\end{tabular}
\end{table}

\subsection{Wavelet-family selection and scale energy}
Haar minimized the validation score in Eq.~\eqref{eq:score} at all three sites. The result reflects the sharp rise and recession of the observed hydrograph: piecewise local contrasts can align closely with the annual pulse. However, smoother families remain scientifically plausible and are retained in the output sensitivity table.

\begin{figure}[H]
\centering
\includegraphics[width=0.98\textwidth]{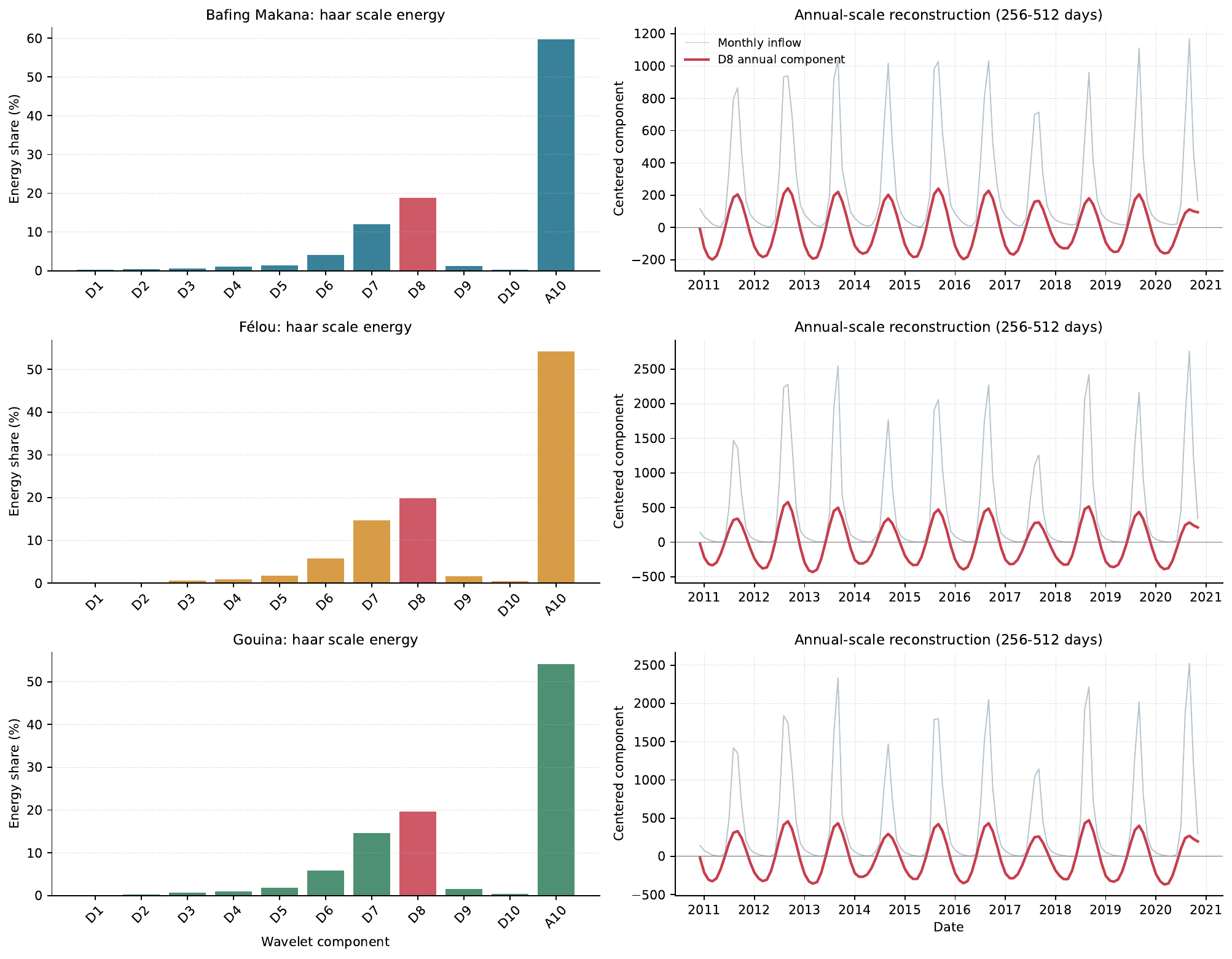}
\caption{Multiresolution energy distribution and reconstructed annual-scale inflow component. Bars show the percentage of total stationary-wavelet-transform energy contained in $D_1$--$D_{10}$ and $A_{10}$ for each site. The reconstructed $D_8$ component, representing approximately 256--512-day variability, is compared with monthly inflow over the final decade to illustrate the timing and amplitude of the annual hydrological cycle.}
\label{fig:energy}
\end{figure}

The annual $D_8$ band accounts for 18.85\%, 19.89\%, and 19.58\% of total SWT energy at Bafing Makana, F\'elou, and Gouina (Table~\ref{tab:energy}; Fig.~\ref{fig:energy}). This component is interpreted as the discrete annual-scale component because its nominal daily period band, 256--512 days, contains the 365.25-day annual cycle. The narrower 300--450-day band used for CWT power and coherence is a continuous-period summary window inside this dyadic interval. The adjacent $D_7$ band contributes 11.98--14.68\%, while $A_{10}$ contributes 54.17--59.74\%. The much smaller $D_1$--$D_4$ shares show that the record's dominant variance is seasonal to interannual rather than purely high-frequency. Importantly, the high $A_{10}$ fraction must not be interpreted as annual seasonality: $A_{10}$ is slower than approximately 1,024 days.

\begin{table}[H]
\centering
\caption{Percentage of total stationary-wavelet-transform energy contained in the selected Haar detail and approximation components. The $D_8$ column identifies the nominal annual-scale band for daily observations, while $A_{10}$ represents variability at periods longer than approximately 1,024 days.}
\label{tab:energy}
\resizebox{\textwidth}{!}{%
\begin{tabular}{lrrrrrrrrrrr}
\toprule
Site & D1 & D2 & D3 & D4 & D5 & D6 & D7 & \textbf{D8} & D9 & D10 & A10\\
\midrule
Bafing Makana & 0.269 & 0.437 & 0.653 & 0.997 & 1.404 & 4.121 & 11.984 & \textbf{18.847} & 1.210 & 0.341 & 59.738\\
F\'elou & 0.069 & 0.236 & 0.535 & 0.950 & 1.712 & 5.708 & 14.675 & \textbf{19.885} & 1.578 & 0.471 & 54.179\\
Gouina & 0.098 & 0.288 & 0.602 & 1.028 & 1.823 & 5.869 & 14.554 & \textbf{19.577} & 1.546 & 0.442 & 54.173\\
\bottomrule
\end{tabular}}
\end{table}

\subsection{Time-varying annual wavelet power}
\begin{figure}[H]
\centering
\includegraphics[width=0.98\textwidth]{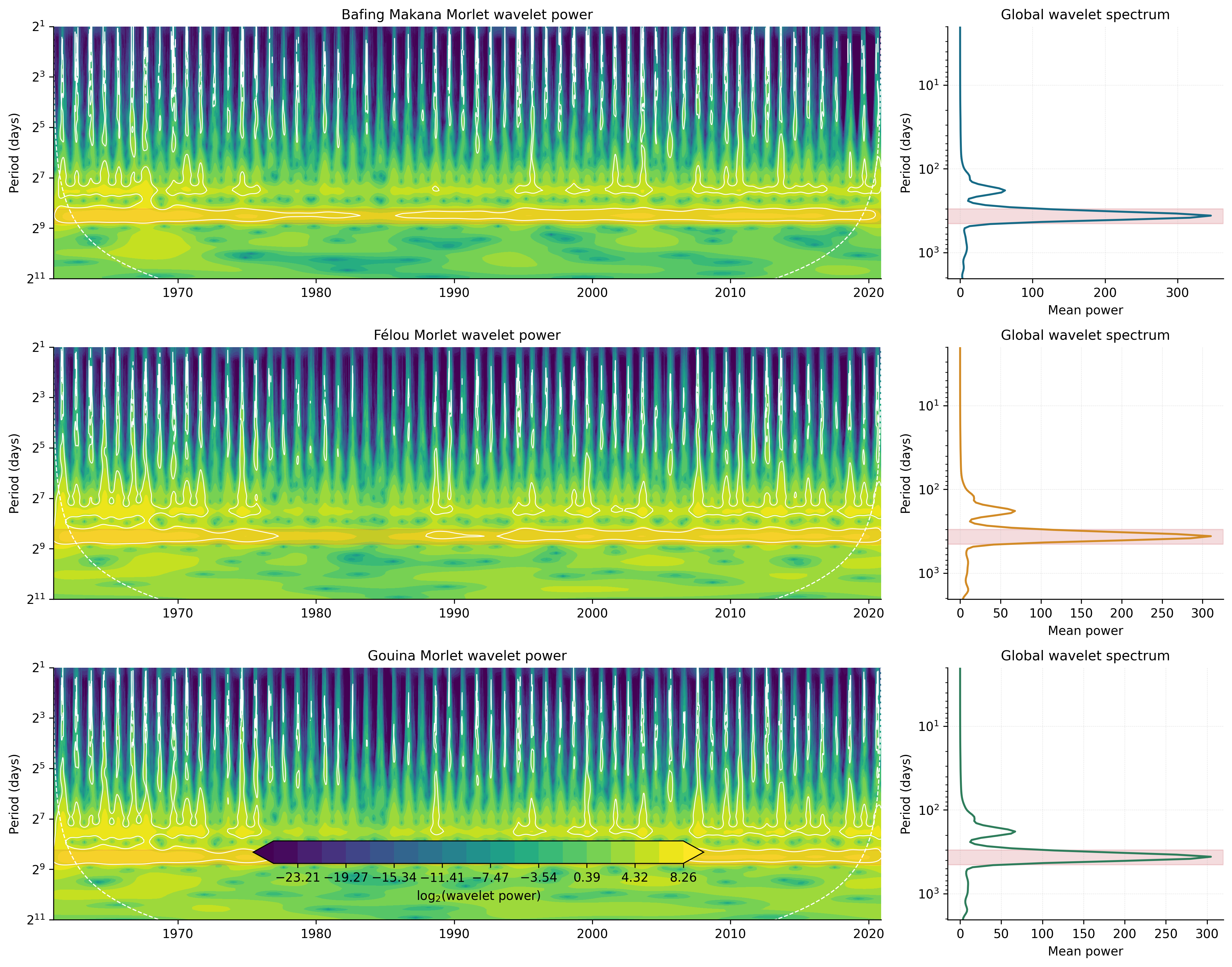}
\caption{Morlet continuous-wavelet power spectra of daily inflow at the three hydropower sites. Colour intensity represents local wavelet power as a function of time and period. White contours delimit regions significant at the 95\% level relative to an AR(1) red-noise background, the curved boundary indicates the cone of influence, and the horizontal reference line marks the annual period.}
\label{fig:cwt}
\end{figure}

Morlet CWT power confirms a persistent annual band, but its intensity varies between years (Fig.~\ref{fig:cwt}). Periods of stronger and weaker annual power are visible, illustrating why a single Fourier amplitude cannot completely describe the series. Interpretation near the record boundaries is restricted by the COI. Mean WASI values are 0.584 (Bafing Makana), 0.542 (F\'elou), and 0.538 (Gouina), with standard deviations of 0.063, 0.074, and 0.076.

\subsection{Annual amplitude, timing, and change points}
\begin{figure}[H]
\centering
\includegraphics[width=0.98\textwidth]{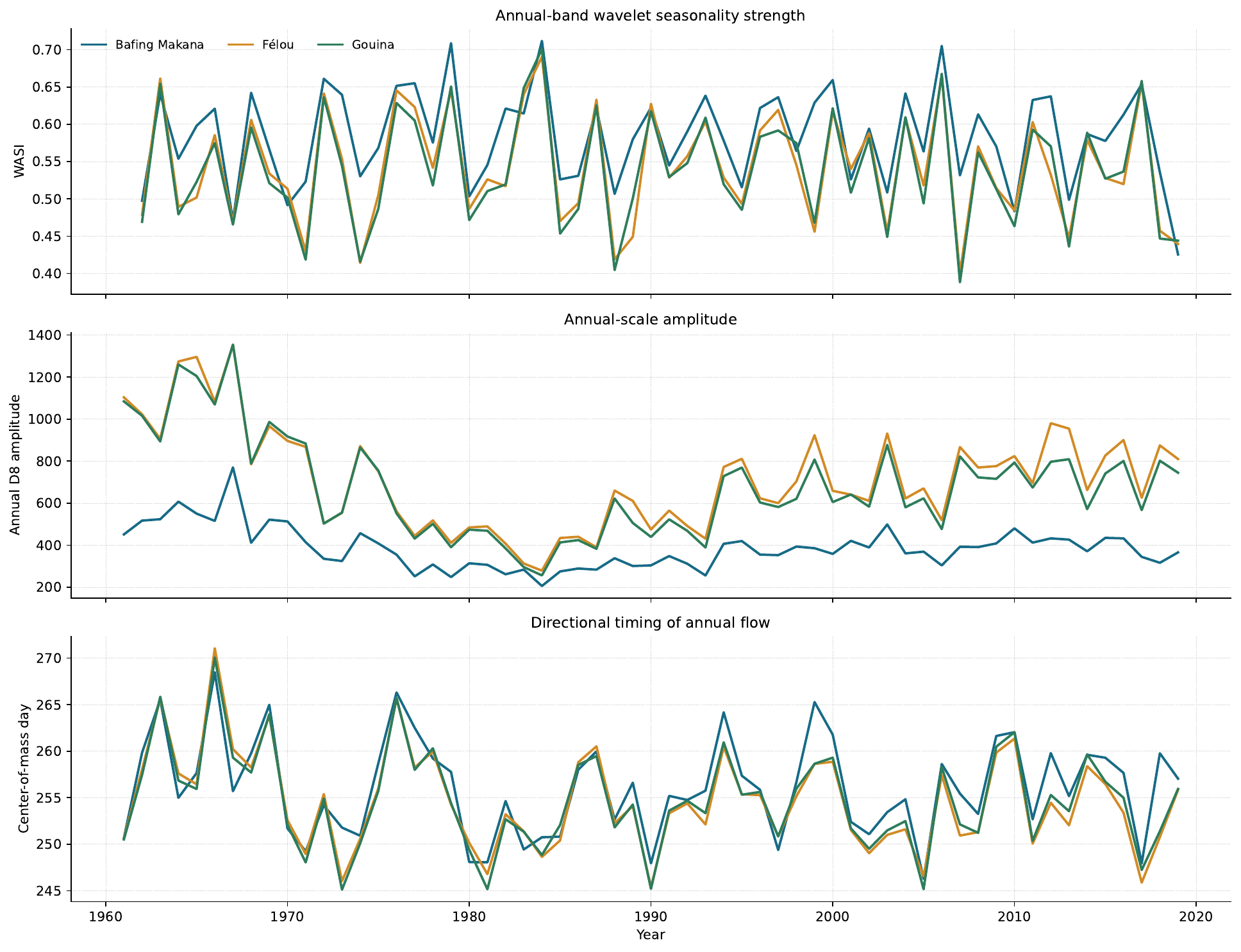}
\caption{Interannual evolution of the Wavelet Annual Seasonality Index, reconstructed $D_8$ amplitude, and annual peak timing. Each value summarizes one hydrological year. Dashed lines represent Sen-slope estimates and are used to assess gradual changes in seasonal strength, magnitude, and phase over the observation period.}
\label{fig:metrics}
\end{figure}

The mean $D_8$ annual-component amplitudes are 386.15 $\mthree$ at Bafing Makana, 716.84 $\mthree$ at F\'elou, and 679.98 $\mthree$ at Gouina. Mean component peaks occur on days 253.75, 252.58, and 252.44, respectively, corresponding to early September. Kendall tests do not identify robust monotonic trends in WASI, annual-component amplitude, or peak day (Fig.~\ref{fig:metrics}). F\'elou's directional concentration showed a nominal increase ($\tau=0.204$, unadjusted $p=0.022$, Sen slope $5.62\times10^{-4}$ year$^{-1}$). However, this result did not remain significant after Bonferroni correction across the 15 site--metric Kendall trend tests ($p_{\mathrm{Bonf}}=0.337$). It is therefore interpreted only as exploratory evidence of possible increased seasonal concentration, not as a robust monotonic trend.

Pettitt tests identify significant shifts in annual-component amplitude or annual mean around 1971--1976. Specifically, amplitude change points are 1975 at Bafing Makana ($p=0.002$), 1971 at F\'elou ($p=0.004$), and 1975 at Gouina ($p=0.001$). These dates overlap the regional hydroclimatic transition associated with the severe Sahel drought era reported in basin studies \citep{Diop2018,Wilcox2018,Bodian2020}. Because several change-point tests were conducted, the Pettitt results were interpreted as a family of related evidence rather than as isolated single tests. The early-1970s amplitude and annual-mean shifts remain the most coherent features of the change-point analysis, whereas monotonic Kendall trends do not survive multiplicity correction. The agreement with known hydroclimatic timing is suggestive, but the test alone cannot separate climate forcing, regulation, rating-curve changes, or data-processing effects.

\subsection{Cross-site dependence}
\begin{figure}[H]
\centering
\includegraphics[width=0.98\textwidth]{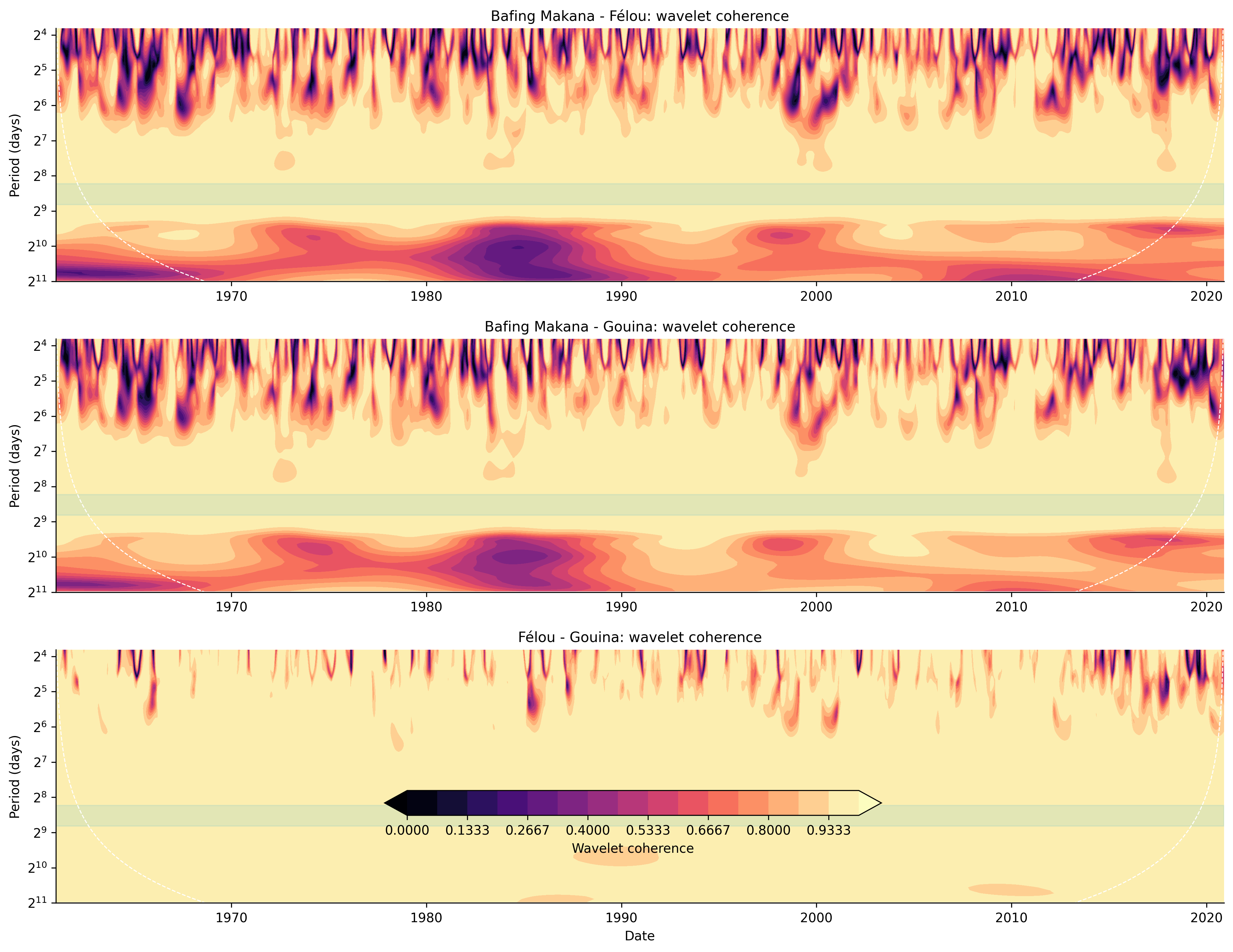}
\caption{Squared wavelet coherence between weekly mean inflows for each pair of sites. Values approaching one indicate strong synchronization at the corresponding time and period. The highlighted 300--450-day band represents annual variability, while coherence at longer periods describes shared interannual behaviour. Results near the boundaries should be interpreted within the cone of influence.}
\label{fig:coherence}
\end{figure}

Raw daily correlations are 0.943 (Bafing Makana--F\'elou), 0.948 (Bafing Makana--Gouina), and 0.994 (F\'elou--Gouina). Correlations of the $D_8$ annual components increase to 0.988, 0.989, and 0.998, respectively. Annual-band mean wavelet coherence is similarly high: 0.988, 0.990, and 0.999 (Table~\ref{tab:coherence}; Fig.~\ref{fig:coherence}). Interannual coherence is lower for pairs involving Bafing Makana (0.703--0.758) but remains high between F\'elou and Gouina (0.975). This pattern indicates a nearly common annual pulse, with more site-specific variability at longer scales upstream versus downstream.

\begin{table}[H]
\centering
\caption{Mean squared wavelet coherence for the three site pairs within the annual and interannual period bands. Values close to one indicate strong time--frequency synchronization of the corresponding inflow signals.}
\label{tab:coherence}
\begin{tabular}{lrr}
\toprule
Pair & Annual (300--450 d) & Interannual (730--2,048 d)\\
\midrule
Bafing Makana--F\'elou & 0.988 & 0.703\\
Bafing Makana--Gouina & 0.990 & 0.758\\
F\'elou--Gouina & 0.999 & 0.975\\
\bottomrule
\end{tabular}
\end{table}

\section{Discussion} \label{sec:discussion}

\subsection{Annual seasonality as a persistent but temporally variable organizing mode}
\label{subsec:discussion_seasonality}

The Fourier, multiresolution, and coherence results consistently identify annual seasonality as the principal organizing feature of the three inflow records. The Fourier harmonic models explain between 73.3\% and 78.7\% of daily variability, the $D_8$ detail contains approximately one fifth of total stationary-wavelet-transform energy, and coherence in the 300--450-day band is close to unity for all site pairs. These results are mutually supportive but describe different aspects of the signal. Fourier regression quantifies the average periodic structure over the complete record, the discrete wavelet decomposition measures how variance is distributed among temporal scales, and continuous-wavelet power identifies when annual variability is locally strong or weak.

The annual cycle is therefore persistent, but it is not stationary in amplitude. The CWT spectra and annual WASI values show pronounced interannual modulation of annual-band power. A year may preserve approximately the same seasonal peak date while exhibiting a substantially different wet-season magnitude, duration, or degree of concentration. This distinction is operationally important: predictability of seasonal timing does not imply predictability of annual water availability. Reservoir planning based only on a mean climatological hydrograph may therefore reproduce the typical timing of inflow while underrepresenting the uncertainty in seasonal volume and peak intensity.

The trend and change-point results further indicate that the long-term evolution of seasonality is not well summarized by a single linear tendency. The only nominally significant Kendall trend, the increase in F\'elou directional concentration, did not survive Bonferroni correction across the 15 trend tests. The corrected results therefore support the conclusion that there is no robust monotonic trend in annual-seasonality strength, reconstructed amplitude, or peak timing over the full record. By contrast, the Pettitt tests identify abrupt changes during the early to mid-1970s. A piecewise-regime interpretation is consequently more appropriate than a uniform-trend interpretation for these records.

This result is hydrologically meaningful. West African hydroclimate has experienced marked multi-decadal transitions, particularly around the severe Sahel drought period, rather than a simple continuous change in one direction \citep{Wilcox2018,Bodian2020,Bruckmann2022,Ndiaye2023}. However, attribution cannot be established from inflow records alone. River inflow reflects the combined effects of precipitation, evapotranspiration, soil-moisture storage, groundwater contribution, catchment modification, hydraulic routing, reservoir regulation, and measurement practices. Without concurrent rainfall, reservoir-release, abstraction, and gauge-history information, the observed change points should be interpreted as statistical evidence of a common regime transition rather than as proof of a uniquely climatic cause.

\subsection{Physical interpretation of the wavelet levels}
\label{subsec:discussion_levels}

A central methodological result concerns the physical interpretation of dyadic wavelet levels. Under the adopted daily sampling convention, the nominal period band associated with detail level $D_j$ is approximately
\[
2^j \leq P < 2^{j+1}\quad\text{days}.
\]
Consequently, $D_4$ represents approximately 16--32-day variability and is associated with submonthly fluctuations rather than annual seasonality. Since
\[
2^8=256 < 365.25 < 512=2^9,
\]
the annual period lies within the nominal $D_8$ band. This provides the mathematical basis for treating $D_8$ as the principal annual-scale component in the present analysis.

The correspondence between a wavelet level and a physical period is nevertheless approximate. The effective frequency response depends on the wavelet family, filter length, boundary treatment, and leakage between neighbouring bands. The annual signal may therefore contribute partly to adjacent components, particularly $D_7$ or $D_9$, even though its central period falls within $D_8$. For this reason, the identification of annual seasonality was not based on dyadic scaling alone; it was also checked against monthly climatology, Fourier spectra, and continuous-wavelet power near 365 days.

This explains why two annual windows appear in the analysis. The SWT component $D_8$ is a broad discrete dyadic reconstruction band, approximately 256--512 days, while the CWT and coherence summaries use a narrower 300--450-day window because continuous wavelets allow a denser period grid. The latter window is centred on the hydrological annual period and remains inside the $D_8$ interval. It is therefore not a conflicting definition; it is a more localized summary of annual power and synchronization within the broader dyadic band.

An energy criterion alone is not an adequate level-selection rule. For example, an approximation such as $A_4$ may retain a large fraction of total variance because it aggregates all variability slower than approximately 16 days. That retained variance includes monthly, seasonal, annual, and interannual components and therefore cannot be labelled annual seasonality. Similarly, the large energy share in $A_{10}$ reflects slow multi-year behaviour rather than the annual cycle. Wavelet level should thus be selected from the target physical period and then validated spectrally and climatologically.

\subsection{Interpretation and robustness of the Haar selection}
\label{subsec:discussion_haar}

The Haar family minimized the validation score in Eq.~\eqref{eq:score} at all three sites. Its compact support and piecewise-constant basis provide an efficient representation of local contrasts, including rapid wet-season onset, pronounced high-flow pulses, and relatively sharp recession phases. The selection is therefore consistent with the asymmetric form of the observed hydrographs.

However, the result should be interpreted as a model-selection outcome under the adopted score, not as evidence that river inflow is physically discontinuous or that Haar is universally superior for hydrological applications. Natural runoff generation, channel routing, and reservoir propagation are continuous processes, even when their sampled hydrographs contain abrupt changes. Smoother wavelets, including Daubechies, Symlet, and Coiflet families, may be preferable when continuity, differentiability, or phase smoothness is important \citep{Daubechies1992,PercivalWalden2000,HadiTombul2018}. This is particularly relevant for derivative-based numerical schemes, Galerkin approximations, data assimilation, or differential-equation representations, where the non-differentiability of the Haar basis can be restrictive.

The sensitivity outputs retained for all candidate families are important in this respect. A scientific conclusion is robust when the main findings---the dominance of the annual band, the timing of the seasonal peak, and the cross-site synchronization---remain qualitatively similar under alternative admissible wavelet bases. The selected family determines the numerical allocation of energy among adjacent levels, but it should not fundamentally change the hydrological interpretation.

\subsection{Implications for MOSSHOOS--Plan4RES scenario generation}
\label{subsec:discussion_mosshoos}

The multiresolution results provide a physically interpretable basis for stochastic inflow scenario construction. Rather than treating daily inflow as a single homogeneous stochastic process, it can be represented as
\begin{equation}
Q_t=L_t+S_t+I_t+H_t,
\label{eq:scale_scenario_decomposition}
\end{equation}
where
\begin{align}
L_t &= A_{10,t}+D_{10,t}+D_{9,t},\\
S_t &= D_{8,t},\\
I_t &= D_{7,t}+D_{6,t},\\
H_t &= \sum_{j=1}^{5}D_{j,t}.
\end{align}
Here, $L_t$ represents slowly evolving interannual and multi-year hydrological conditions, $S_t$ represents the dominant annual pulse, $I_t$ captures seasonal-transition and intraseasonal variability, and $H_t$ contains shorter-duration fluctuations. These categories are operational interpretations rather than perfectly separated physical processes, because hydrological events may project onto several neighbouring scales.

The decomposition should not be used as an automatic denoising rule. A small coefficient is not necessarily measurement error. Short-scale details can encode physically meaningful flood initiation, rainfall--runoff pulses, wet-season onset, and recession irregularities. Removing such coefficients without an independent error model may produce scenarios that are visually smooth but hydrologically unrealistic, particularly in the tails. Thresholding should therefore be justified using known measurement uncertainty and evaluated through out-of-sample tests of temporal dependence, quantiles, annual volume, and extreme-event behaviour.

The strong annual-band coherence also has direct implications for multivariate modelling. Generating the three sites independently would fail to reproduce the observed regional synchronization and could create unrealistic scenarios in which one site experiences an anomalously strong wet season while the others remain dry. A more defensible representation is
\begin{equation}
S_{p,t}=\lambda_pF_t+\varepsilon_{p,t},
\label{eq:common_annual_factor}
\end{equation}
where $F_t$ is a common annual-scale factor, $\lambda_p$ is the site-specific loading, and $\varepsilon_{p,t}$ is a residual component. At interannual scales, the weaker coherence involving Bafing Makana implies that site-specific residual variability must be retained rather than imposing perfect dependence at all frequencies.

These multivariate scenarios can be used as inputs to the Plan4RES seasonal-storage model or to MOSSHOOS SDDP and HJB extensions. In an SDDP formulation, slow wavelet components may contribute to the Markov hydrological state, while annual and shorter-scale components define conditional stagewise inflow distributions. In a continuous-time HJB formulation, the low-frequency state may influence the time-dependent mean-reversion level or volatility of inflow. In both cases, preserving cross-scale and cross-site dependence is preferable to introducing independent additive noise.

\paragraph{River continuity and lagged dependence}
The high annual-band coherence between the three sites confirms strong upstream--downstream synchronization, particularly between F\'elou and Gouina. However, the present analysis does not explicitly estimate hydrological transfer times along the river network. A natural extension is to compute lagged dependence between sites, for example
\begin{equation}
\rho_{pr}(\ell)=\operatorname{Corr}\left(Q_{p,t},Q_{r,t+\ell}\right),
\qquad
\ell^*_{pr}=\operatorname*{arg\,max}_{\ell\in[-L,L]}\rho_{pr}(\ell),
\end{equation}
or the corresponding lagged correlation of annual and intraseasonal wavelet components. Such an analysis could help separate basin-wide seasonal forcing from routing delays and operational regulation. It is left as a perspective because robust physical interpretation requires verified site coordinates, river-distance information, reservoir-operation data, and possibly daily or event-scale routing models. The present coherence results should therefore be interpreted as evidence of synchronization, not as an explicit travel-time model.

\subsection{Limitations and scope of inference}
\label{subsec:discussion_limitations}

Several limitations define the scope of the present conclusions. First, the input file does not contain observation-status codes, gauge-quality flags, rating-curve history, or detailed provenance fields. The statistical workflow is fully reproducible from the available records, but the analysis cannot distinguish physical observations from possible coding, censoring, or rating-curve effects where the input file provides no metadata. This is why recorded zeros are retained as observed values and not reclassified as missing values in the principal analysis.

Second, the coordinates used in Fig.~\ref{fig:map} are not analytical variables; they provide only geographical orientation. All numerical results are derived from the date--site--inflow records. Third, the present analysis is descriptive rather than attributive. The association between change points and regional hydroclimatic transitions cannot distinguish among climatic forcing, catchment alteration, regulation, abstraction, or measurement changes. Attribution would require additional meteorological and operational covariates.

Fourth, continuous-wavelet significance and wavelet coherence are affected by serial dependence, finite-record length, smoothing, and boundary effects. Results near the cone of influence were therefore interpreted cautiously, and significance was assessed relative to an AR(1) red-noise background \citep{MaraunKurths2004,MaraunEtAl2007,RodriguezMurillo2020}. Coherence indicates shared time--frequency behaviour and does not establish causal direction.

Fifth, multiple trend, change-point, and site-pair tests increase the probability of isolated significant findings. The nominal increase in F\'elou directional concentration did not remain significant after Bonferroni correction across the 15 Kendall trend tests and is treated as exploratory. Finally, the wavelet-family score uses the Fourier seasonal representation as one validation target. This favours agreement with the average periodic cycle and may not select the optimal family for forecasting, extreme-event preservation, or dynamical-system approximation. Future evaluation should include blocked out-of-sample prediction, tail fidelity, autocorrelation preservation, and sensitivity to alternative boundary rules.

\section{Conclusions}\label{sec:conclusions}

This study provides a reproducible multiscale characterization of nearly 60 years of daily inflow records, spanning January 1961 to November 2020, at Bafing Makana, F\'elou, and Gouina. The combined climatological, Fourier, discrete-wavelet, and continuous-wavelet results show that the three sites are governed by a pronounced and highly synchronized annual hydrological cycle, with mean peak timing concentrated in early September. At daily resolution, the annual period is represented primarily by the nominal $D_8$ band, corresponding approximately to 256--512 days, rather than by a level-4 approximation.

Under the adopted validation criterion, the Haar wavelet provided the best-performing decomposition, while the $D_8$ component accounted for approximately 19\% of total stationary-wavelet-transform energy at each site. Fourier harmonic models explained between 73\% and 79\% of daily inflow variability and effectively described the average seasonal cycle, whereas the Morlet continuous-wavelet transform revealed substantial temporal variation in annual-band power that could not be represented by a stationary harmonic model.

No robust monotonic trend was identified in annual-seasonality strength, reconstructed amplitude, or peak timing after multiplicity correction. However, statistically significant change points in annual-component amplitude or mean inflow were detected during the early to mid-1970s. These transitions are consistent with known regional hydroclimatic disturbances, although their attribution cannot be established without complementary rainfall, reservoir-operation, and gauge-history information.

The near-unity annual-band coherence among the three sites demonstrates that their seasonal inflows should not be generated independently in stochastic applications. Future scenario models should preserve a common annual factor while allowing site-specific and scale-dependent variability, particularly at interannual and shorter time scales. The main hydrological finding is basin-specific, whereas the validation protocol combining Fourier analysis, dyadic wavelet reconstruction, CWT power, and coherence is transferable to other West African basins with long daily records. The resulting framework establishes a transparent hydrological basis for multivariate wavelet-informed inflow generation and its subsequent integration into MOSSHOOS--Plan4RES seasonal-storage and stochastic-optimization models.

\section*{Data and code availability}
The analysis code, Google Colab notebook, environment file, derived CSV tables, and publication figures are available in the accompanying reproducibility package. The observed daily inflow file was used within the MOSSHOOS project workflow. Public redistribution of the raw inflow records is subject to authorization by the responsible data owner. If redistribution of the raw file is not permitted, the authors will provide the code and derived outputs and indicate the procedure by which qualified researchers may request access to the original inflow data.

\section*{Author contributions}
\textbf{Steeven Belvinos Affognon:} Conceptualization, methodology, software, formal analysis, visualization, writing--original draft. \textbf{Babacar Mbaye Ndiaye:} Supervision, hydropower interpretation, stochastic-optimization framing, writing--review and editing. \textbf{Pierre Mendy:} Wavelet methodology, mathematical review, writing--review and editing. \textbf{Cheikh M. F. Kebe:} Energy-system interpretation, project supervision, writing--review and editing.

\section*{Declaration of competing interest}
The authors declare that they have no known competing financial interests or personal relationships that could have appeared to influence the work reported in this paper.

\section*{Funding}
This research benefited from the support of the Fondation Math\'ematique Jacques Hadamard (FMJH) through the Programme Gaspard Monge pour l'Optimisation, la recherche op\'erationnelle et leurs interactions avec les sciences des donn\'ees (PGMO). 


\section*{Acknowledgements}
The authors acknowledge the MOSSHOOS project partners and the institutions responsible for the hydrological records. The authors also thank colleagues who provided comments on data provenance, coordinate verification, multiplicity correction, and the distinction between discrete and continuous annual-period bands.

\nocite{*}
\bibliographystyle{elsarticle-harv}
\bibliography{wavelet_seasonality_references}
\end{document}